\documentclass[prl,aps,twocolumn,showpacs,superscriptaddress]{revtex4}
\usepackage{amsmath}
\usepackage{graphicx}
\usepackage{dcolumn}

\begin{document}

\title{Guided nucleation of superconductivity on a graded magnetic substrate}

\author{M. V. Milo\v{s}evi\'{c}}
\affiliation{Departement Fysica, Universiteit Antwerpen,
Groenenborgerlaan 171, B-2020 Antwerpen, Belgium}

\author{W. Gillijns}
\affiliation{INPAC - Institute for Nanoscale Physics and Chemistry,
Nanoscale Superconductivity and Magnetism \& Pulsed Fields Group,
Katholieke Universiteit Leuven, Celestijnenlaan 200 D, B-3001
Leuven, Belgium}

\author{A. V. Silhanek}
\affiliation{INPAC - Institute for Nanoscale Physics and Chemistry,
Nanoscale Superconductivity and Magnetism \& Pulsed Fields Group,
Katholieke Universiteit Leuven, Celestijnenlaan 200 D, B-3001
Leuven, Belgium}

\author{A. Lib\'{a}l}
\affiliation{Departement Fysica, Universiteit Antwerpen,
Groenenborgerlaan 171, B-2020 Antwerpen, Belgium}

\author{V. V. Moshchalkov}
\affiliation{INPAC - Institute for Nanoscale Physics and Chemistry,
Nanoscale Superconductivity and Magnetism \& Pulsed Fields Group,
Katholieke Universiteit Leuven, Celestijnenlaan 200 D, B-3001
Leuven, Belgium}

\author{F. M. Peeters}
\affiliation{Departement Fysica, Universiteit Antwerpen,
Groenenborgerlaan 171, B-2020 Antwerpen, Belgium}
\date{\today}

\begin{abstract}
We demonstrate the controlled spatial nucleation of
superconductivity in a thin film deposited on periodic arrays of
ferromagnetic dots with gradually increasing diameter. The
perpendicular magnetization of the dots induces vortex-antivortex
molecules in the sample, with the number of (anti)vortices
increasing with magnet size. The resulting gradient of antivortex
density between the dots predetermines local nucleation of
superconductivity in the sample as a function of the applied
external field and temperature. In addition, the compensation
between the applied magnetic field and the antivortices results in
an unprecedented enhancement of the critical temperature.
\end{abstract}

\pacs{73.23.-b, 74.78.Na}

\maketitle

The enhancement of the superconducting critical parameters is
arguably the most prominent objective of mesoscopic
superconductivity. For example, it is already well established that
critical magnetic fields rise in submicron samples to values
substantially larger than those found in bulk materials \cite{1},
and both critical field and current can be enhanced by
nanoengineered periodic \cite{2} and quasi-periodic pinning
\cite{3}.

The best current method of preventing the deterioration of the
superconducting state under external magnetic fields consists of
locally counteracting the applied field by an array of ferromagnets
with perpendicular anisotropy. These field-compensation effects can
lead to a substantial increase of the $T_c(H)$ boundary in a field
range determined by the magnetization of the dots \cite{5}. More
precisely, the maximum of the $T_c(H)$ boundary shifts away from
zero field, to a value determined by the number of {\it
antivortices} in the sample \cite{6}. Namely, if magnets are capable
of generating vortex-antivortex molecules in the superconductor
\cite{7}, an applied field that provides exactly the same number of
vortices will lead to a complete vortex-antivortex annihilation
between the dots, which will maximize $T_c(H)$.
\begin{figure}[b]
\includegraphics[width=\linewidth]{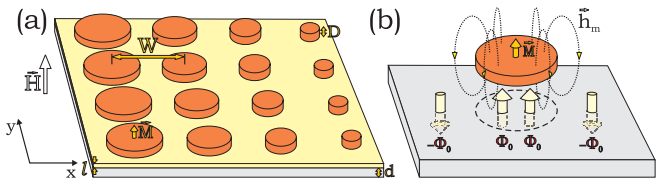}
\caption{\label{Fig1}(color online) (a) A square lattice of magnetic
dots with linearly increasing size on top of a superconducting film
and an oxide layer. (b) The stray field of each dot ($\vec{h}_m$)
may induce vortex-antivortex molecules in the superconductor
\cite{7}.}
\end{figure}

In an attempt to achieve the ultimate field compensation, here we
study a thin superconducting film evaporated on top of a square
array of out-of-plane magnetized dots {\it of variable size}, i.e.
parallel periodic rows of dots increasing in diameter [see Fig.
\ref{Fig1}(a)]. We expect that such gradual magnetic texturing will
enhance superconductivity in different parts of the sample at
different applied fields, and therefore {\it should lift the
$T_c(H)$ characteristics as a whole}. Since vortex-antivortex (V-Av)
physics is fundamentally important in such a system where different
dots induce different numbers of V-Av pairs [see Fig.
\ref{Fig1}(b)], we also investigate the resulting structure of V-Av
complexes.

Our sample is a 50 nm thick Al film with critical temperature
$T_{c0}=1.350$ K. Experimentally estimated values of coherence
length and penetration depth are $\xi(T$=0)=117 nm and
$\lambda(0)$=98 nm, respectively. A $5$ nm thick Si insulating layer
separates the Al film from the underlying square array of circular
magnetic dots with $W=2~\mu$m lattice spacing [see Fig.
\ref{Fig1}(a)], so that any proximity effect can be neglected. The
ferromagnetic dots have diameters of $0.2$, $0.4$, $0.6$, $0.8$ and
$1.0\mu$m and consist of a $2.5$ nm Pt buffer layer covered with a
[$0.4$ nm Co/$1.0$ nm Pt]$_{10}$ multilayer with magnetization
perpendicular to the sample surface. Our numerical simulations are
performed both within the Ginzburg-Landau (GL) formalism and by
molecular dynamics (MD) using vortex-vortex and magnetic dot-vortex
interaction potentials derived from London theory \cite{8}. For thin
superconducting films (always effectively type-II) the results of
these two theories should converge.

\paragraph{Vortex-antivortex complexes}
To analyze theoretically the vortex-antivortex configuration and
their influence on the S/N phase boundary, we employ the GL
theory. The two GL equations for the order parameter $\Psi$ and
the vector potential ${\bf A}$ are solved self-consistently, as
detailed in Ref. \cite{7}. The ground-state vortex configurations
are determined by comparing the Gibbs free energy
$\mathcal{F/F}_0=V^{-1}\int (2({\bf A}-{\bf A}_{0}){\bf
j}-|\Psi|^{4})d{\bf r}$ of all found stable vortex states, where
$\mathcal{F}_0=H_c^2\big/8\pi$, and all quantities are
dimensionless. Here, ${\bf j}$ denotes the supercurrent density,
$A_0$ the applied vector potential, and $V$ the sample volume. In
what follows, most of the calculated quantities will be expressed
(back) in real units for direct comparison with experiment.
\begin{figure}[t]
\includegraphics[width=\linewidth]{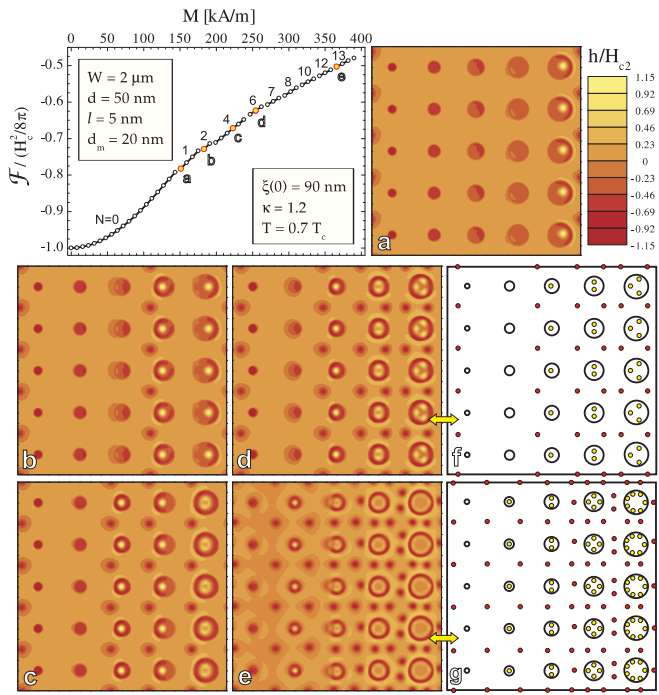}
\caption{\label{Fig2}(color online) The GL free energy of the
superconducting state as a function of the magnetization of the
magnetic dots (field-cooled regime). (a-e) Local magnetic field
distribution in the sample for representative vortex-antivortex
configurations ($N=1,2,4,6,13$ antivortices per row respectively),
and comparison with MD simulations (f,g).}
\end{figure}

Fig. \ref{Fig2} shows the calculated evolution of the ground-state
energy of the superconducting film as a function of the
magnetization of the magnetic dots, for other parameters taken as in
the experiment ($d$, $l$, and $d_m$ denote the thickness of the
film, oxide and the magnetic dots respectively). As schematically
shown in Fig. \ref{Fig1}(b), each dot generates an inhomogeneous
stray field which favors the nucleation of vortex-antivortex (V-Av)
pairs. For a given magnetization, V-Av pairs first nucleate under
the largest dots, due to their larger magnetic moment [see Fig.
\ref{Fig1}(a)]. In interaction with other dots, antivortices occupy
interstitial positions, while vortices remain under the largest dots
at a somewhat off-center position. With increasing magnetization,
more V-Av pairs are induced in the sample, and they gradually appear
around smaller dots as well [see Figs. \ref{Fig2}(b-d)]. Increased
density of antivortices between the magnetic dots leads to their
`crystallization' into an interstitial lattice [Fig. \ref{Fig2}(e)].
Results of the GL theory are nicely corroborated by MD simulations
[see Figs. \ref{Fig2}(f,g)].

Two important conclusions follow from the above results. First, our
gradiated magnetic texture leads to a gradient of antivortex density
along the sample (for vortices as well, but they are localized under
the magnetic dots). Second, depending on the spacing between the
magnetic dots and the number of nucleated antivortices, some of them
can penetrate even the regions free of nucleated vortex-antivortex
pairs [see area around smallest dots in Fig. \ref{Fig2}(e)]. This
shows that antivortices can detach from the original magnetic dot
and move towards other dots, guided by an overall magnetic
potential.

\begin{figure}[b]
\includegraphics[width=\linewidth]{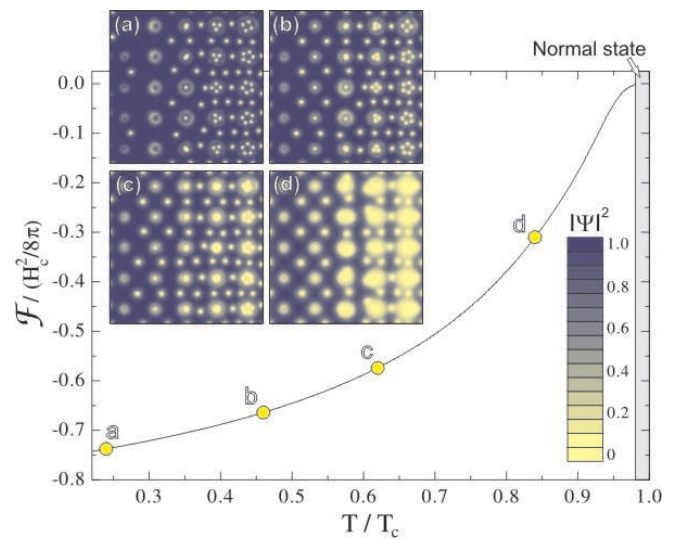}
\caption{\label{Fig3}(color online) The evolution of the Cooper-pair
condensate (and its energy) under the gradiated array of magnetic
dots as a function of temperature (for $M=300$ kA/m).}
\end{figure}
The latter conclusions are even more evident if realistic structural
defects are introduced in the calculation. In Fig. \ref{Fig3}(a) we
show the vortex-antivortex state obtained for $M=300$ kA/m, where we
allowed for minor deviations of the size of the magnets ($<1$ nm)
compared to the experimentally suggested values. As a result, some
dots with the same nominal size generate different number of V-Av
pairs in the sample and consequently the overall magnetic profile
becomes less uniform and antivortices penetrate easier the areas
around the dots of insufficient moment for the nucleation of V-Av
pairs. Interestingly, the magnetic profile in the sample changes
considerably with increasing temperature [see Figs.
\ref{Fig3}(b-d)]. Fig. \ref{Fig3} shows that higher temperature
leads to: (i) appearance of new V-Av pairs around some dots; (ii)
merging of confined vortices under the dots into a giant-vortex
\cite{9}; (iii) increased rigidity of the antivortex lattice between
the magnetic dots; and (iv) gradual destruction of
superconductivity, starting from areas around the largest dots.
Molecular dynamics simulations disregard the suppression of
superconductivity under the magnetic dots and do not take into
account the large vortex cores - as a consequence they fail to
reproduce the GL results at temperatures above $0.8~T_c$.

\paragraph{Controlled nucleation of superconductivity}
The above findings indicate that with increasing temperature
superconductivity is longest preserved in the region around the
smallest magnetic dots. However, it is already a well-known fact
that nanostructuring of superconducting films by perpendicularly
magnetized dots leads to a shift of the $H-T$ phase boundary, a
phenomenon known as field-induced-superconductivity \cite{5}. This
phenomenon is asymmetric with applied external field, i.e. fields
parallel to the moment of magnetic dots enhance superconductivity
(due to stray field compensation and vortex-antivortex annihilation
between the dots) while applied antiparallel field only suppresses
superconductivity further. As a result, the $H-T$ boundary shifts
with respect to $H=0$, and exhibits maximal $T_c(H)$ at a field
determined by the magnetic moment of the dots. In our present sample
however different rows of magnetic dots have different magnetic
moments, and it is by no means obvious how the $H-T$ boundary will
look like.

\begin{figure}[t]
\includegraphics[width=\linewidth]{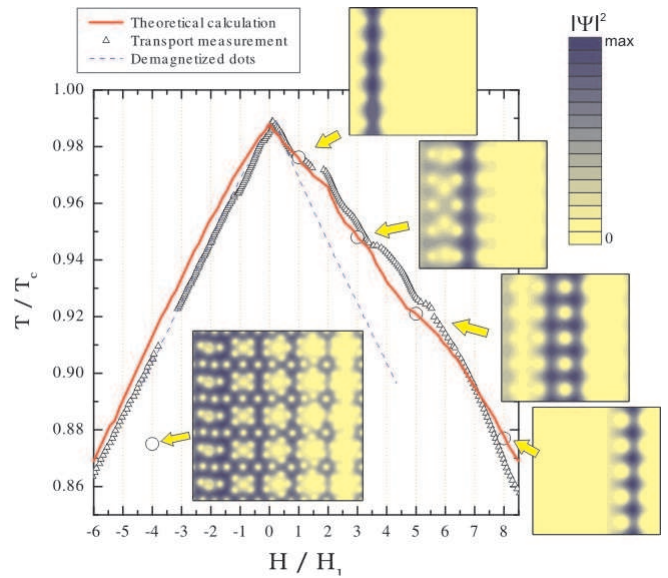}
\caption{\label{Fig4}(color online) The superconducting/normal (S/N)
$H-T$ phase boundary obtained by transport measurements, compared to
the case of demagnetized dots and the theoretical data (for $M=450$
kA/m, $\lambda(0)=100$ nm, $\xi(0)=110$ nm). The insets show the
Cooper-pair density plots at (H,T) values denoted by open dots.}
\end{figure}

The results of our transport measurements and GL simulations for the
superconducting/normal phase $H-T$ boundary are shown in Fig.
\ref{Fig4}. In the theoretical simulations, the nucleation of
superconductivity was monitored locally, i.e. maximal
$|\Psi|^2>10^{-12}$ was declared as a superconducting state. In the
case of negative applied field, we obtained an identical phase
boundary for both magnetized and demagnetized dots (dashed curve in
Fig. \ref{Fig4}). In this case, since superconductivity always
nucleates along the line between the rows of the two smallest dots,
$T_c(H)$ shows an approximately linear behavior. Deeper inside the
superconducting phase (lower $T$) the applied magnetic field induces
additional antivortices in the sample, increasing the
configurational complexity. As an illustration, the inset of Fig.
\ref{Fig4} at $T=0.86~T_c$ and $H=H_{-4}$, i.e. matching field
adding 4 antivortices per magnetic dot, shows the resulting
structure of both antivortex molecules around the magnetic dots and
an antivortex lattice between them.

On the other hand, for applied positive magnetic field we observe a
clear enhancement of superconductivity in our sample. As detailed in
Ref. \cite{6}, the highest $T_c$ in magnetically structured
superconducting films is always obtained at the applied field
matching the number of antivortices per periodic unit cell of the
sample. Since our sample contains magnetic dots of five different
sizes and correspondingly different number of induced antivortices
around them ($N$=1+2+2+5+8), we achieve the latter condition for
maximal $T_c$ at several values of applied magnetic field! As a
consequence the overall maximum of $T_c$ remains at $H=0$, but the
$T_c(H)$ boundary is lifted tremendously as compared to the
reference sample. The superposition of compensation phenomena in
different parts of the sample leads to several cusps in $T_c(H)$
\footnote{Resembling commensurability effects, see Ref. \cite{2}.},
but the general trend is clear - as the applied magnetic field is
increased, superconductivity nucleates first in the regions of the
sample with matching antivortex density. As illustrated in Fig.
\ref{Fig4}, the nucleation line shifts from smallest (at $H=H_1$)
towards largest dots (for $H=H_8$). Such guided nucleation of
superconductivity is not only a unique fundamental phenomenon, but
it is of value for potential multi-channel applications, and, as
shown in Fig. \ref{Fig4}, leads to an unprecedented enhancement of
superconductivity in an applied magnetic field.

To summarize, thanks to the specific design of the arrays of
magnetic dots with varied size, and the resulting gradient in
antivortex density in the superconducting film, we achieved the
{\it highest} overall critical temperature of superconducting
films in the presence of applied magnetic field. We also show that
nanoengineered arrays of magnetic dots enable the control of local
nucleation of superconductivity by simply changing the {\it
magnitude} of applied magnetic field.

This work was supported by the Flemish Science Foundation (FWO-Vl),
the Belgian Science Policy (IAP), and the ESF-NES program. W.G.,
A.V.S. and A.L. acknowledge individual support from FWO-Vlaanderen.

\end{document}